\documentclass[journal]{IEEEtran}
\usepackage{graphicx}
\usepackage{amsmath, amsfonts}
\usepackage{epsfig}
\usepackage{subfigure}

\usepackage{cite}

\usepackage{color}

\usepackage{multirow}
\usepackage{rotating}

\usepackage{arydshln}

\usepackage{epstopdf}

\begin{document}

\title{Simplifying Karnaugh Maps by Making Groups of a Non-Power-of-Two Number of Elements}

%
%

\author{Mario~Garrido
	\thanks{The author is with the Department of Electronic Engineering, ETSI de Telecomunicaci\'on, Universidad Polit\'ecnica de Madrid, 28040 Madrid, Spain, e-mail: mario.garrido@upm.es} 
	\thanks{This work was supported by the Ram\'on y Cajal Fellowship RYC2018-025384-I of the Spanish Ministry of Science, Innovation and Universities.}
}	
	

\maketitle

\begin{abstract}
	When we study the Karnaugh map in the switching theory course, we learn that the ones in the map must be combined in groups of $a \times b$ elements, being $a$ and $b$ powers of two. The result is the logic function described as a sum of products. This paper shows that we can also make groups where $a$ and/or $b$ are equal to three. This does not result in a sum of products, but in a logic function that is simpler than the sum of products in terms of logic gates. This idea is extended later in the paper to groups of $2^n-1$ elements.
\end{abstract}


\section{Introduction}

When we study the Karnaugh map~\cite{Hol05,Tin95}, we learn that the ones in the map must be combined in groups of $a \times b$ elements, where $a$ and $b$ are powers of two~\cite{Hol05}. Other shapes and rectangles of other sizes are not allowed. A simple example is when there are three ones in a row of the Karnaugh map, as shown in Fig.~\ref{fig:1x3}(a). In this case, two groups of two elements are created. One with the first element and the middle one and the other group with the middle element and the last one. The alternative shown in Fig.~\ref{fig:1x3}(b), where three elements are grouped together, is not allowed.

When the rule for grouping elements is followed, the result is a logic function represented as a sum of products (SOP). This representation has the advantage that it leads to a circuit with low delay~\cite{Ber19}. By contrast, the resulting circuit is generally not efficient in terms of the number of logic gates. For this reason, it is reasonable to wonder why we look for a SOP representation if it does not lead to hardware-efficient results. This question should make us reconsider the motivation to use the Karnaugh map. As achieving a hardware-efficient circuit is a main goal when designing a digital circuit, we should figure out ways to provide it, such as current optimization methods to simplify boolean functions~\cite{Tin95,Ber19,Ban09,Ber13}. Conversely, achieving low delay, as the Karnaugh map actually does, is a usually a secondary goal in most digital designs. In this sense, the aim of the current paper is to provide a new perspective to the Karnaugh map that makes it suitable for the design of hardware-efficient circuits, not only low delay ones. This is done by a new way to group the ones in the Karnaugh map that considers groups of a number of elements that is not a power of two. This widens the understanding of the Karnaugh map and makes it a more powerful tool.

The proposed approach is developed in the paper as follows. First, making groups of three, six and nine elements is studied in Section~\ref{sec:Gr3}. Then, the ideas are generalized to groups of $2^n-1$ elements in Section~\ref{sec:Gr2nm1}, which completes the approach. Later, in Section~\ref{sec:HowTo} it is explained how to use the proposed approach to improve the explanation of the Karnaugh map. Finally, the main conclusions of the paper are summarized in Section~\ref{sec:Conclusions}.

\section{Making groups of three, six and nine elements} 
\label{sec:Gr3}

\begin{figure}[t]
	\centering{\includegraphics[width=6.5cm]{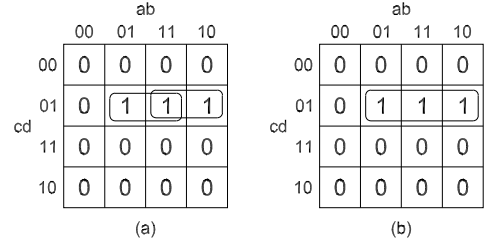}}
	\caption{Grouping three elements. (a) Conventional approach. (b) Proposed approach.\label{fig:1x3}}
\end{figure}

Figure~\ref{fig:1x3}(a) shows how to group three ones in a Karnaugh map according to the conventional approach, which consist of making two groups of two elements. The logic function for this case is 
\begin{equation}\label{eq:1}
	f = a\overline{c}d + b\overline{c}d,
\end{equation}
which can be implemented with 5 2-input logic gates. As a general criterion throughout the paper, the number of logic gates is counted as the number of 2-input logic gates.

The alternative presented in Fig.~\ref{fig:1x3}(b) groups all the three elements together. In this case, the second row corresponds to $\overline{c}d$, whereas the three last columns correspond to the function $a+b$. This leads to
\begin{equation}\label{eq:2}
	f = (a+b)\overline{c}d, 
\end{equation}
which can be implemented with 3 logic gates. Therefore, grouping three elements together is more hardware-efficient than making two groups of two elements.

\begin{figure}[t]
	\centering{\includegraphics[width=3cm]{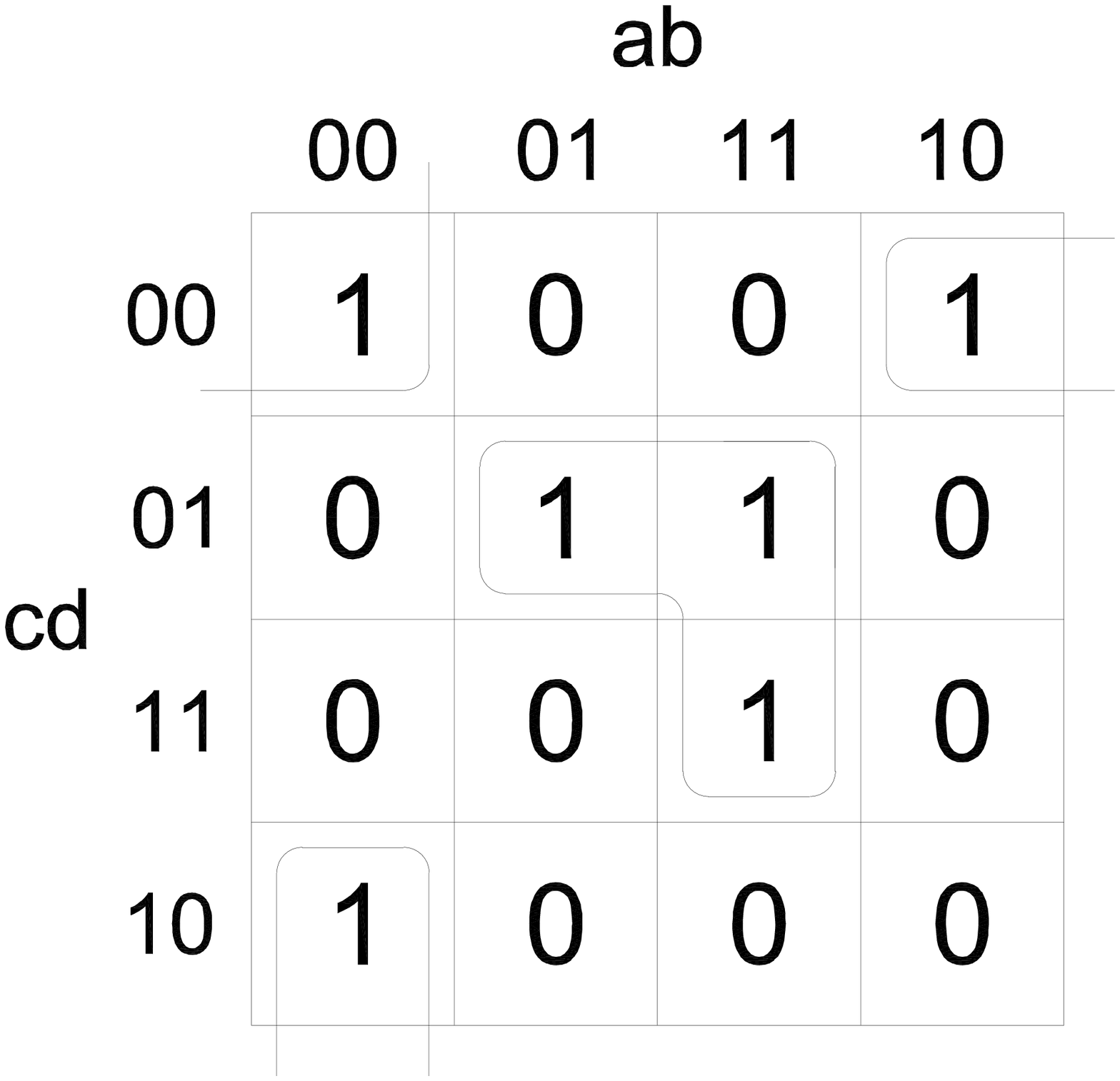}}
	\caption{Example where groups of three elements are made and these groups have an $L$ shape.\label{fig:1x3L}}
\end{figure}

This idea can be extended to groups of three elements that form an $L$ shape, as is shown in Fig.~\ref{fig:1x3L}. The three ones in the center correspond to $bd(\overline{c}+a)$ and those in the corners correspond to $\overline{b} \;\overline{d}(\overline{c}+\overline{a})$.

\begin{figure}[t]
	\centering{\includegraphics[width=6.5cm]{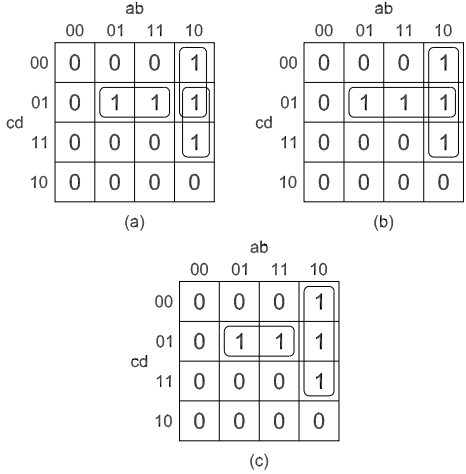}}
	\caption{Two groups of three elements. (a) Conventional approach. (b) Making groups of three elements. (c) Proposed approach.\label{fig:2x3}}
\end{figure}

When there are two groups of three elements that intersect as in Fig.~\ref{fig:2x3}, the conventional approach in Fig.~\ref{fig:2x3}(a) obtains the logic function by making three groups of two elements, which results in
\begin{equation}
	f = b\overline{c}d + a\overline{b}\overline{c} + a\overline{b}d.
\end{equation}
The circuit used to calculate this equation requires 8 logic gates. By contrast, making groups of three elements as in Fig.~\ref{fig:2x3}(b) results in
\begin{equation}
	f = (a+b)\overline{c}d + (\overline{c}+d)a\overline{b},
\end{equation}
which requires 7 logic gates. However, there is an even better alternative shown in Fig.~\ref{fig:2x3}(c), which consists of a group of two elements and a group of three elements and leads to
\begin{equation}
	f = b\overline{c}d + (\overline{c}+d)a\overline{b},
\end{equation}
and requires 6 logic gates. 

The examples in Figs.~\ref{fig:1x3} and~\ref{fig:2x3} lead to two interesting conclusions. First, a group of three elements is more hardware-efficient than two groups of two elements. Second, a group of two elements is more hardware-efficient than a group of three elements. These conclusions serve as decision rule when making the groups.

\begin{figure}[t]
	\centering{\includegraphics[width=6.5cm]{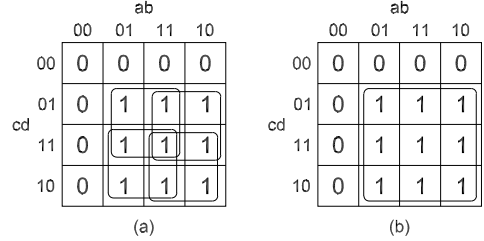}}
	\caption{Grouping a square of three by three elements. (a) Conventional approach. (b) Proposed approach.\label{fig:3x3}}
\end{figure}

Another interesting case is when there is a square of $3 \times 3 $ ones, as shown in Fig.~\ref{fig:3x3}. The use of the conventional approach requires to make four squares of $2 \times 2$ and leads to
\begin{equation}
	f = bd + ad + bc + ac,
\end{equation}
which requires 7 logic gates.

The alternative for this case is to group all the 9 elements together. The three last columns correspond to the function $a+b$ and the three last rows to $c+d$, which results in
\begin{equation}
	f = (a+b)(c+d).
\end{equation}
In this case, the calculation only requires 3 logic gates, which is a significant reduction of the hardware cost with respect of grouping the elements in squares of $ 2\times 2$.

\begin{figure}[t]
	\centering{\includegraphics[width=6.5cm]{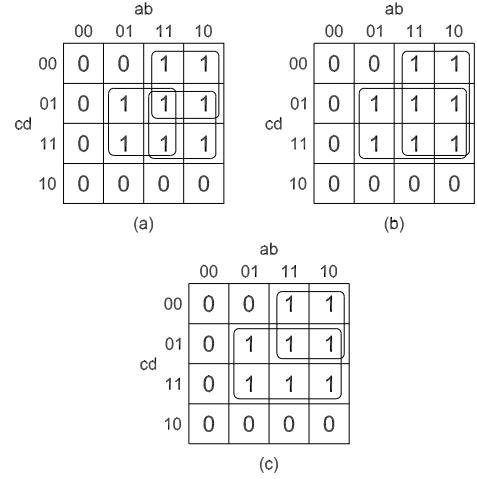}}
	\caption{Making groups of six elements. (a) Conventional approach. (b) Using only groups of six elements. (c) Proposed approach.\label{fig:6}}
\end{figure}

As a final example of making groups with a number of elements that is multiple of three, Fig.~\ref{fig:6} highlights the case of grouping 6 elements together. The conventional approach solves the Karnaugh map in Fig.~\ref{fig:6}(a) by making three groups of $2 \times 2$, and obtains
\begin{equation}
	f = a\overline{c} + bd + ad,
\end{equation}
which requires 5 logic gates.

The alternative of using two groups of 6 elements in Fig.~\ref{fig:6}(b) results in
\begin{equation}
	f = a(\overline{c}+d) + (a+b)d,
\end{equation}
which also requires 5 logic gates.

Finally, by using a group of 6 elements and a group of 4 elements as in Fig.~\ref{fig:6}(c) the resulting logic function is
\begin{equation}
	f = a\overline{c} + (a+b)d.
\end{equation}
In this case, the number of logic gates is reduced to 4.

This example illustrates the facts that a group of 4 elements is more hardware-efficient than a group of 6 elements, whereas a group of 6 elements is more hardware-efficient than two groups of 4 elements.

\section{Making groups of $2^n-1$ elements}
\label{sec:Gr2nm1}

The ideas for groups with a number of elements that is a multiple of three can be generalized to groups of $2^n-1$ elements where $n \in \mathbb{N}$. These elements must be embedded in a rectangle of size $2^i \times 2^j$ where both $i, j \in \mathbb{N}$ and $i+j = n$. According to this, there will be a single element in the $2^i \times 2^j$ rectangle that is not a one. This element will be excluded from the group by using an OR function.

\begin{figure}[t]
	\centering{\includegraphics[width=6.5cm]{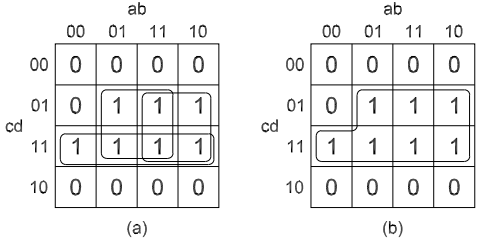}}
	\caption{Grouping seven elements. (a) Conventional approach. (b) Proposed approach.\label{fig:7}}
\end{figure}

As an example, Fig.~\ref{fig:7} shows a group of 7 elements. In this case, $i = 1$, $j = 2$, $n=3$, $2^i \times 2^j = 2^n = 8$ and $2^n-1 =7$. According to the conventional approach in Fig.~\ref{fig:7}(a), the ones are grouped in three groups of four elements, leading to
\begin{equation}
	f = cd + bd + ad.
\end{equation}
This logic function requires 5 logic gates.

According to the proposed approach, the ones are embedded in a rectangle of $2 \times 4$ whose logic function is $d$. Inside the rectangle, the function is $(a+b+c)$. This leads to 
\begin{equation}
	f = d(a+b+c),
\end{equation}
which results in 3 logic gates. Again, this strategy reduces the number of logic gates.

\section{Making the Karnaugh map a more powerful tool}
\label{sec:HowTo}

The proposed approach allows for presenting the Karnaugh map as an optimization tool with two possible goals: To reduce the delay of the circuit or to obtain a hardware-efficient digital circuit.

In order to reduce the delay of the circuit, we derive the SOP expression with the conventional approach by using the following rules:
\begin{itemize}
	\item Group the ones in the Karnaugh map in squares or rectangles of $2^i \times 2^j$ elements.
	\item Borders of the Karnaugh map are connected to the opposite borders, which allows to connect elements from both extremes.
	\item A one in the map may be included in ore or several groups.
	\item Each group must include at least a one that is not included in any other group. Otherwise, the group is redundant.
	\item Groups must be made with the aim of making the smallest number of groups and include the largest number of ones in these groups. 
\end{itemize}

In order to obtain a hardware-efficient circuit, we incorporate the ideas presented in this paper and consider making groups of a number of elements that is not a power of two. This transforms the design rules into:
\begin{itemize}
	\item Group the ones in the Karnaugh map in squares or rectangles of $a \times b$ elements where $a, b \in {1, \ldots, 4}$, or groups of $2^n-1$ elements embedded in a square or rectangle of size $2^i \times 2^j$, being $i+j = n$.
	\item Borders of the Karnaugh map are connected to the opposite borders, which allows to connect elements from both extremes.
	\item A one in the map may be included in ore or several groups.
	\item Each group must include at least a one that is not included in any other group. Otherwise, the group is redundant.
	\item Groups must be made with the aim of making the smallest number of groups and include the largest number of ones in these groups. However, a group of 2 ones is preferable to a group of 3 ones where the first or last one is already included in another group. Likewise, a group of 4 ones is preferable to a group of 6 if the 2 ones of difference are already included in another group.
\end{itemize}

\section{Conclusion}
\label{sec:Conclusions}

In this paper, a new way to understand the Karnaugh map has been presented. The new approach enables groups of ones whose size is not a power of two, which is not allowed in the conventional approach. As a result, the new approach allows for a further simplification of the logic functions, leading to digital circuits with smaller number of gates.

This enriches the explanation of the Karnaugh map, which can be explained as a tool to either minimize the delay of the circuit or reduce the number of logic gates.

%

\bibliographystyle{IEEEtran}
\bibliography{latexBib}

\vspace{0.5cm}

\textbf{Mario Garrido} (M'07-SM'19) received the M.Sc. degree in electrical engineering and the Ph.D. degree from the Technical University of Madrid (UPM), Madrid, Spain, in 2004 and 2009, respectively. In 2010 he moved to Sweden to work as a postdoctoral researcher at the Department of Electrical Engineering at Link\"oping University. From 2012 to 2019 he was Associate Professor at the same department. In 2019 he returned to UPM, where he holds a Ram\'on y Cajal Research Fellowship. 
	

\end{document}